\documentclass{aastex}                             %

\newcommand{\beq}{\begin{equation}}
\newcommand{\eeq}{\end{equation}}
\begin{document}
\title{Magnetic flux captured by an accretion disk}
\author{H.~C.\ Spruit}
\affil{Max-Planck-Institut f\"ur Astrophysik,  Box 1317, 85741, Garching, Germany}
\email{ henk@mpa-Garching.mpg.de}
\author{Dmitri~A.~Uzdensky}
\affil{Princeton University, Department of Astrophysics, Peyton Hall, Princeton, NJ 08544}
\email{ uzdensky@astro.princeton.edu}


\begin{abstract}
We suggest a possible mechanism of efficient transport of the large-scale external magnetic flux inward through a turbulent accretion disk. The outward drift by turbulent diffusion which limits the effectiveness of this process  is reduced if the external flux passes through the disk in concentrated patches. Angular momentum loss by a magnetocentrifugal wind from these patches of strong field adds to the inward drift. We propose that magnetic flux  accumulating in this way at the center of the disk provides the `second parameter' determining X-ray spectral states of black hole binaries, and the presence of relativistic outflows.

\end{abstract}


{\tt\obeylines
}



\section{Introduction}
\label{sec-intro}

Magnetic fields in thin accretion disks can be of two conceptually different 
kinds. On the one hand, there are the chaotic, small-scale (compared with the 
disk radius) magnetic fields generated by the dynamo action and field-line 
stretching in the turbulent shear flow (Shakura \& Sunyaev 1973; Stone et al. 
1995; Brandenburg et al.  1996; Machida et al. 2000; Sano et al. 2004). 
These are the fields now believed to produce the effective viscous stress 
responsible for the accretion flow in most types of accretion disks (Balbus 
\& Hawley 1991). On the other hand, one might think of fields captured from 
the environment. If such `trapped' large scale fields can be dragged in with 
the accretion flow, they would be amplified to large strengths on their way 
to the center of the disk, because of the decreasing disk area (e.g., 
Bisnovatyi-Kogan \& Ruzmaikin~1976).

Though both kinds of fields can be useful for interpreting a range of phenomena
in accretion disks, there is an important conceptual difference between them. 
Locally generated turbulent fields can appear and disappear in situ. Like 
hydrodynamic shear turbulence, their average strength, length scales and 
time scales would reflect the local conditions in the disk. The statistical 
properties of these fields will therefore be predictable functions of the 
disk conditions, in particular the local accretion rate.

This is not the case for trapped fields. The net vertical magnetic flux through the disk surface is conserved, not changeable by local processes in the disk (as a consequence of ${\rm div}~ \bf B=0$). It can change only by  advection into or diffusion out of the disk through its outer edge. If magnetic flux can be trapped at all, it therefore constitutes an {\em additional parameter}, or degree of freedom characterizing the disk, and one would expect this property to show up in observations.


\section{Trapping ordered fields in disks}
\label{sec-problem}

\subsection{Indications of strong ordered fields}
\label{subsec-evidence}

The X-ray emission from neutron star and black hole binaries can be
classified into `soft' and `hard' states, understood as originating 
from a standard optically thick, cool accretion disk and an optically 
thin Comptonizing `cloud', respectively. The hard states are typically 
associated with lower luminosities. The correlation between accretion 
rate (as inferred from total X-ray luminosity) and the spectral state 
of the source is not very direct, however. The so-called very high state 
in black hole transients, for example, is a hard state at high luminosity 
(prototypically seen in Nova Muscae, Ebisawa et al. 1994). In the persistent 
black hole binary GRS 1915 (Belloni et al. 2000) spectral state changes take 
place as loops of varying size in hardness-intensity or color-color diagrams. 
In the persistent neutron star accreters, there is generally a close relation 
between intensity and hardness of the X-rays, taken as indicative of an 
underlying dependence of these parameters on the accretion rate. This 
relation, however is not stable: day-to-day shifts are seen in these 
relations (e.g., Homan et al. 2002; Di Salvo et al. 2003; Schnerr et al. 2003).

These variations in the relation the between spectrum and the (inferred) accretion rate are hard to understand in a conventional accretion disk scenario (with or without magnetic coronas), since this scenario is too deterministic. Apart from statistical fluctuations due to turbulence (e.g.\ Gilfanov et al. 1999), the local state of the accretion flow, in the inner region where the X-ray luminosity is produced, is governed by just one parameter: the instantaneous accretion rate in this region.

These observations are much easier to understand if there is a ``second 
parameter'' influencing the accretion flow. There are few plausible 
candidates for such a parameter. Trapped magnetic flux is one, and, 
as we shall argue here, it has ideal properties. For example, it can 
plausibly vary on time scales of days or longer, corresponding to the 
time for accretion from the outer edge of the disk. At the same time 
shorter time scales are also possible, corresponding to internal 
instabilities of the trapped field further inside the disk.

A second, again indirect, indication for magnetic field trapping is the
presence of relativistic jets, seen from both black hole and neutron star
accreters (e.g. Cir X-1). They are present in many objects, but usually 
not all the time (a noticeable exception being the steady jet in SS433). 
The mechanism of choice for producing these is the magneto-centrifugal 
mechanism (e.g.\ Bisnovatyi-Kogan \& Ruzmaikin 1976; Blandford \& Payne 
1981; for a tutorial see Spruit 1996) for which a strong open magnetic 
field has 
to be present above the disk. The field due to magnetic disk turbulence would
produce a magnetized corona, but the small length scale of the turbulence (of the order of the disk thickness) will strongly limit the height to which the field extends above the disk, thereby also limiting the possible acceleration of an outflow. A large scale, ordered field such as that provided by trapped flux would certainly work better. Indeed, almost all calculations of magnetic outflows have in fact assumed such configurations.

A third consideration is the collimation of jets. Even magnetic jet acceleration requires an additional agent to focus the outflow, especially if the opening angle is to be as small as a few degrees. The toroidal field associated with a rotating magnetic outflow has some collimating effect, but only for modest degrees of collimation.  For the high degree of collimation often seen, internal (kink) instabilities are particularly effective at destroying the toroidal field component that does the collimating, and an additional mechanism is needed (Eichler 1993, Spruit et al. 1997). An external dense medium, such as the pre-SN envelope in the collapsar model for GRB can have the right properties. In many cases, however, there is no visible evidence of such a collimating environment, especially in the case of the microquasars. The only plausible alternative left then is collimation by an ordered poloidal magnetic field anchored in the disk (`poloidal collimation'). As shown in Spruit et al. (1997), close to perfect collimation is possible for a range of parameters of such a poloidal field, as long as the size of the disk is large enough compared with the source region of the jet. 
This condition is satisfied for X-ray binaries and AGN disks, but not for 
the disks in cataclysmic variables, where the ratio of outer to inner disk 
radius is of the order of a factor 100 or less. In fact, no jets are known 
from CVs, though there is ample evidence for uncollimated outflows from 
these systems.


\subsection{The difficulty of trapping flux in a disk}
\label{subsec-difficulty}

The virtues of trapped magnetic flux as described above are somewhat offset 
by a theoretical obstacle. The small scale processes that give rise to the 
exchange of angular momentum allowing mass to accrete through a disk (most 
likely magnetic turbulence, Shakura \& Sunyaev 1973; Hawley et al. 1996) may 
also act like an effective magnetic diffusivity. As magnetic field lines are 
dragged in with the accretion flow, they diffuse out by the same process that 
causes the accretion. As noted first by van Ballegooijen (1989; see also 
Heyvaerts et al. 1996; Agapitou \& Papaloizou 1996; Livio et al. 1999) this 
limits the effectiveness of field line `dragging' to much lower values than 
one might at first sight expect. This is because the dragging causes a kink 
in the field line direction at the disk surface. As a result, the length 
scale relevant for the diffusion of the field is the disk thickness $H$ 
rather than the radial scale $r$ of the field. The outward diffusion speed 
is set by reconnection of the radial field component across the disk, not 
the radial diffusion of the vertical component. If the turbulent diffusivity 
for the magnetic field is similar to the turbulent viscosity (i.e., the 
effective magnetic Prandtl number of order unity), this limits the angle 
of the field lines with respect to the vertical to a value of order $H/r$, 
and dragging is inefficient.

A way around this obstacle is to argue that the magnetic Prandtl number could 
be significantly smaller than unity. The small scale motions associated with 
the exchange of angular momentum act in the $r-\phi$ plane [in cylindrical 
coordinates $(r,\phi,z)$], whereas the diffusion of the radial field component 
across the disk is effected by exchanges in the $(r,z)$ plane. A suitable form 
of anisotropy of the turbulence might lower the effective magnetic Prandtl 
number. Numerical simulations of magnetic disk turbulence in which the relevant
components of the diffusivity were monitored with passively advected vector 
fields, however, have not shown any significant anisotropy (Brandenburg and 
Spruit 1998, unpublished). The diffusive properties of magnetic turbulence in 
disks appear to have received little attention in numerical simulations so far;
we believe a closer look at this problem would be valuable.

We note as an aside, that, while the kink in the field at the disk surface 
causes a radial outward force (by the curvature of the field lines), this 
force is not what prevents the field from being dragged in with the accretion 
flow. The limiting effect is entirely kinematic, due to diffusion of the 
magnetic field in the assumed turbulence. A displacement by the radial force 
causes the angular momentum to deviate from its surroundings; as a result 
radial forces are almost instantaneously balanced by a small deviation from 
Keplerian rotation. A small radial drift comes in only as a result of drag 
associated with the difference in orbital velocity with respect to the 
surroundings (see section~\ref{subsec-outward-drift} below).

These considerations assume that the magnetic field is weak enough that 
the forces it exerts can be neglected compared to gravity. This looks 
like a perfectly reasonable assumption, since the main idea of creating 
a strong magnetic field in the inner regions is to start with a weak field 
in the outer regions of the disk. However, this is correct only if the field 
that is being dragged in is in a diffuse form. If it is intermittent and 
concentrated into small patches of high field strength, the forces it exerts 
can not be neglected, and the situation is different.

While the accretion flow alone is thus not a very good way of trapping 
magnetic flux we find that other, quite effective mechanisms exist.


\section{A better flux trap}
\label{sec-model}
 
In this section we describe a possible mechanism for transport of 
large-scale vertical magnetic flux inward across a turbulent accretion 
disk. A key element in our proposal is the realization that, in order 
to understand how magnetic flux moves around in such a disk, one  
has to look beyond the concept of `turbulent magnetic diffusivity'. 
Of particular importance is the phenomenon of turbulent diamagnetism, 
i.e., the property of turbulence to expel magnetic field from regions 
of strong fluid motions and to concentrate it in small patches at the 
boundaries of turbulent cells. As a result, the magnetic field in a 
turbulent medium is generally expected to be strongly intermittent
(e.g., Vishniac~1995; Schekochihin et al. 2004).
As we shall see later in this section, the spatial intermittency of 
the magnetic field should play a very important role in the overall 
transport of magnetic flux.%
\footnote
{This effect has also been found to be important in the interstellar
medium turbulence (e.g., Lazarian \& Vishniac 1996).}
In fact, in our model we will take this 
notion to the limit by assuming that the plasma in the disk is segregated 
into two coexisting phases: most of the disk area is covered by a usual 
turbulent medium, whereas there is also a second phase comprised of 
small patches of very strong magnetic field that suppresses turbulence. 
We start with a brief review of these effects.


\subsection{Magnetic fields in turbulence}
\label{subsec-magnetic-turbulence}

The interaction of an externally imposed magnetic field with a turbulent 
flow has been a subject of investigation since the 1960's, especially in 
connection with observations of the Sun, where it is responsible for the 
very patchy pattern that is so characteristic of magnetic fields at the 
its surface. These studies, as well as numerical magnetohydrodynamic (MHD) 
simulations, have led to a rather detailed theoretical picture for the 
advection of magnetic fields  by a turbulent medium.


\subsubsection{Flux expulsion}
\label{subsubsec-flux-expulsion}

An image of the solar surface shows the magnetic field to be very fragmented: 
on all length scales the magnetic field is concentrated into flux bundles of 
strong field, separated by convective cells with little magnetic flux. In the 
magnetic regions, the field is strong enough to suppress convection. In a 
sunspot this is evident through the low amplitudes of its internal velocity 
field and its reduced temperature. On the other hand, the convective flow is 
observed to quickly concentrate any newly emerging magnetic flux into the 
vertices between convective cells (esp. with the high resolution observations 
now available, e.g., Lites et al. 2004; Berger et al. 2004). This `flux 
expulsion' effect of convective turbulence has been understood theoretically 
since the early days of astrophysical MHD (Zeldovich 1956; Parker 1963).

Numerical simulations (Weiss 1966; Proctor \& Weiss 1982) show that the 
expulsion process takes place in two stages.  A pattern of overturning 
cells concentrates the flux of an initially weak magnetic field into its 
vertices, on a time scale of the order of one overturning time of the cells. 
This is initially a kinematic effect: the flux is just advected passively by 
the flow. The increasing strength of the field thus wound up then reacts back 
on the flow, suppressing the flow speeds in the regions of concentrated flux. 
In the interior of the cells, the mixture of fields of opposite directions 
caused by the overturning motion decays by Ohmic diffusion on a longer time 
scale. At the end of the process, a nearly complete separation has taken 
place between the turbulent, field-free cells, and magnetic flux patches 
with suppressed flow.

An important aspect of the flux expulsion process is that it has to happen 
only once: when a flux bundle is formed by the process described, it tends 
to remain in concentrated form by its continued interaction with the 
surrounding fluid flow. It can fragment, and flux can be redistributed 
between neighboring bundles (processes observed abundantly on the Sun), 
but its field strength remains high compared with the that in the surrounding 
flow.


\subsubsection{Analogy with superconductors}
\label{subsubsec-superconductors}

The separation between flow and magnetic flux has an interesting analogy with 
superconductivity. While magnetic flux is excluded from a superconductor, the 
material (a type I superconductor, say) is normally conducting again above a 
critical field strength. If boundary conditions are arranged such that a given 
fixed amount of magnetic flux passes through the material, cooling below the 
critical temperature causes a separation into superconducting regions without 
magnetic flux and normally conducting regions into which the magnetic flux is 
concentrated. The precise nature of the non-superconducting regions depends 
on the type of superconductor (1 vs 2). In the case of magnetic flux in a 
turbulent flow, the turbulence plays the role of superconductivity, expelling 
the flux. The suppression of turbulence by a strong magnetic field is analogous to the lifting of superconductivity by a magnetic field.


\subsection{Behavior of concentrated fields}

\begin{figure}
\plotone{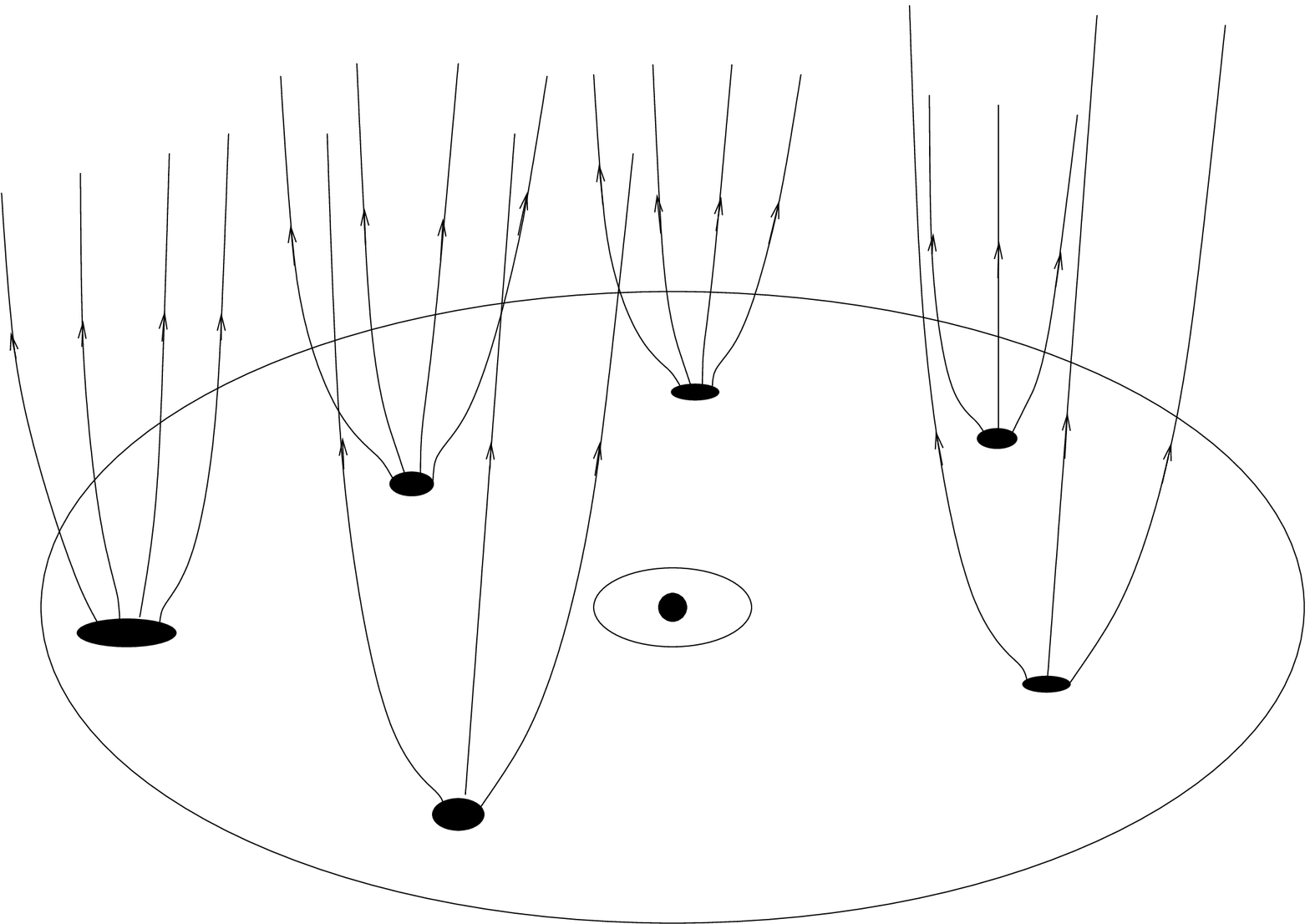}
\caption{Magnetohydrodynamical turbulence in the accretion disk 
leads to the concentration of the weak external large-scale magnetic 
field threading the disk into small patches of strong field 
(shown by black ovals). These magnetized patches rapidly drift 
inward because of the enhanced angular momentum loss caused by
magneto-centrifugally driven wind.\label{fig-1}}
\end{figure}

We now assume that a large portion of the net vertical flux threading a 
disk gets concentrated into patches of strong field (see Fig.~\ref{fig-1}). 
This has two important effects: first, it reduces reconnection across the 
disk, and second, the magnetic patches can lose angular momentum effectively. 
The first effect reduces the outward diffusion, the second causes the field 
to drift inward.

As is the case with magnetic fields in convective turbulence, we assume now 
that the field in an accretion disk locally becomes strong enough to suppress 
the turbulence generated by the magneto-rotational instability (MRI). Some 
evidence for this behavior appears to have been seen already in numerical 
simulations of magnetic disk turbulence (Machida~et~al. 2000). 

The concentrated fields will have especially noticeable effects near the 
surface of the disk. Because of the magnetic pressure, the gas density 
inside the patch is reduced. The resulting buoyancy causes the field to 
remain nearly vertical at the surface (again as seen on the Sun). 

If only the dynamic interaction with the surrounding flow is important, 
the strength of the field becomes of the order of equipartition with the 
turbulent kinetic energy of the flow: just high enough for the field to 
avoid most of the tangling and overturning effect of the flow. 

The field strength can increase further by two effects. Near the surface, radiative cooling can reduce the internal pressure. To the extent that heat transport in the disk is carried by turbulence, its reduction inside the magnetic patch implies a lower heat flux, causing the surface to be cooler inside it. By vertical hydrostatic balance, this reduces the internal pressure and hence increases the field strength. This effect will probably be limited mostly to regions where the disk is convective, as in the case of the Sun, where the reduction of convective heat transport causes sunspots to be cool. The effect could be strong in the outer regions of the disk most likely to be convective, hence of particular importance for the trapping of flux from an external field.

Secondly, mass loss through a wind will occur on just those field lines that connect from the disk surface to the external field. The resulting `emptying' of the field lines also lowers the internal pressure in the field, and hence increases the field strength.


\subsection{Reduced diffusion}

As in the case of the Sun and the numerical simulations mentioned, 
the flux stays in concentrated form once patches are formed by 
interaction with the turbulence. The outward diffusion of trapped 
fields by turbulent reconnection across the disk that is seen in 
kinematic (advection + diffusion) calculations is then reduced once 
the magnetic field locally becomes strong enough to avoid being tangled. 
The remaining diffusion takes place not by physical displacement by the 
fluid motions in the magnetic turbulence, but through reconnection in 
the atmosphere of the disk (see \S~\ref{subsubsec-recn}).


\subsection{Enhanced angular momentum loss in patches}
\label{subsec-enhan}

In addition to the reduced diffusion of concentrated fields, it is quite 
likely that the patches will lose angular momentum rather effectively, 
leading to an additional inward drift. 

The field of a patch, vertical near the surface, fans out above the disk 
into a more horizontal field (the `canopy' in solar physics terminology). 
At the (radially) outward-facing side of a patch, this horizontal field 
has the right direction and strength to effectively accelerate the plasma  
by the magneto-centrifugal mechanism (e.g., Bisnovatyi-Kogan \& Ruzmaikin 
1976; Blandford \& Payne 1982). Whereas the mean field would be insufficiently 
strong to cause much of a centrifugal flow, the concentrated patches will 
almost certainly be effective. 

Angular momentum loss by a magnetic wind involves two steps. First, 
conditions at the disk surface have to be right to launch a mass flux 
on a field line. This requires a minimum field strength and a minimum 
outward inclination away from the vertical (Blandford \& Payne 1982; 
Ogilvie \& Livio 2001). Secondly, in order to carry a wind, a field 
line also has to be open, i.e., it has to extend to a sufficient distance 
from the disk instead of connecting back to some other location on the disk 
surface. 

The first condition (launching) limits the angular momentum loss of a patch 
to its radially outward side. On the inward side the field lines have the 
wrong inclination. The second condition means that the angular momentum loss 
will take place just on those field lines that connect the external field to 
the disk. The foot points of the external field will change continually by 
reconnection against the turbulent disk field, but at any point in time 
conditions for launching will be satisfied on a significant fraction of them. 
The regions contributing to angular momentum loss are therefore not identical 
to the strong-field patches themselves. In order not to complicate the 
nomenclature in the following, however, we will use the term `patch' 
interchangeably for an area of concentrated field and for the parts 
of them where angular momentum loss takes place. 

Thus, we see a large number of small cool islands of strong magnetic field 
drifting through the stormy sea of the turbulent disk (Fig.~\ref{fig-1}). 
Even though the total covering fraction of these magnetized patches is 
small, they can carry a large fraction of the vertical magnetic flux. 
Because the magnetic torque is not linear but quadratic in the field 
strength, the angular momentum is removed from these patches much faster 
than from the rest of the disk. As a result, the patches accrete faster 
than the bulk of the disk matter. But what's important here is the fact 
that it is precisely these same patches that carry a large portion of 
the overall net vertical flux; therefore, the rapid inward motion of 
the patches leads to a very efficient transport of the flux. 
Note that it is not necessary for the patches to maintain their identity 
for more than an orbital time scale in order for the processes described 
to be effective.

Suppose that the disk is threaded by some large-scale vertical field with 
average flux density (i.e., the average field strength)~$B_0$. We assume 
that this average field is relatively weak so that it cannot suppress the 
MRI in the disk. This means that
\beq
\beta_0 \equiv {{8\pi P}\over{B_0^2}} \gg 1 \, ,
\label{eq-def-beta0}
\eeq
where $P$ is the disk gas pressure (in this paper we assume that 
gas pressure dominates over the radiation pressure inside the disk). 
Because of this, strong turbulence develops almost everywhere in the disk. 
This turbulence, in turn, leads to the concentration of the large scale 
magnetic flux into a large number of small sparsely distributed patches. 
The field in these patches increases until it is strong enough to locally 
suppress the~MRI. Thus we can estimate the magnetic field strength $B_p 
\ll B_0$ inside the patches as
\beq
B_p\simeq \sqrt{{8\pi P}\over{\beta_{\rm MRI}}} \, ,
\label{eq-Bp}
\eeq
where $\beta_{\rm MRI}=O(1)$ is the minimal plasma-beta required for MRI 
to be unstable. In our simple model we assume that the vertical magnetic 
field $B_p$ is uniform inside each patch and is the same in all patches. 
Under the assumption that most of the overall vertical magnetic flux 
threads the disk by going through the patches, we can express the surface 
covering fraction $f$ of the patches as
\beq
f= {B_0\over{B_p}} \simeq \sqrt{\beta_{\rm MRI}\over{\beta_0}} \ll 1 \, .
\label{eq-def-f}
\eeq

Because of the strong magnetic field in the patches, the gas pressure inside 
them is lower than the ambient gas pressure. Indeed, from the horizontal 
pressure balance between a patch and the surrounding disk it follows that 
the patch gas pressure~$P_p$ is
\beq
P_p=P+P_{\rm turb}-{B_p^2\over{8\pi}} = P(1+\alpha-\beta_{\rm MRI}^{-1}) \, ,
\label{eq-Pp}
\eeq
where we have parametrized the turbulent pressure (both dynamic and magnetic) 
in the disk as $P_{\rm turb} = \alpha P$. Thus we see that $P_p$ may differ 
significantly from the outside pressure.


\subsection{Inward drift}
\label{subsec-inward-drift}

We can now address the question of the radial transport of vertical
magnetic flux by the inward radial drift of magnetized patches.
A strong magnetic field is very efficient in removing angular 
momentum from the patches by magnetic braking and/or through a 
thermally or magneto-centrifugally driven wind. The magnetic 
torque exerted on a patch of a surface area $a^2$ is
\beq
\tau_{\rm patch} =|{1\over{4\pi}}\, 2r a^2 \, B_p B_{\phi,p}|{\gamma\over{2\pi}}\, r a^2 B_p^2 \, ,
\label{eq-tau_patch}
\eeq
where we have parametrized the strength of the toroidal magnetic field 
$B_{\phi,p}$ just above (and below) the surface of the patch as
\beq
|B_{\phi,p}| = \gamma B_p \, .
\label{eq-Bphi}
\eeq
The value of $\gamma$ depends on the mass flux in the wind.
We discuss this dependence quantitatively in Appendix~1 using 
some well-known properties of cold magnetocentrifugal wind solutions. 
From equation~\ref{eq-fmu} it is seen that $\gamma$ can be larger or
smaller than unity. Values larger than unity correspond to cases of
large mass flux, for which the Alfv\'en surface is close to the base of
the outflow, and the azimuthal field component becomes substantial already 
close to the base (Spruit 1996, Anderson et al. 2004). The angular momentum
flux from the patch will be dominated by those field lines (the outward
facing ones) where mass-launching conditions are favorable and $\gamma$
is large.

The magnetic torque (\ref{eq-tau_patch}) leads to an inward drift of 
the patch with a radial velocity $-v_{\rm dr}$ which can be estimated
by equating $\tau_{\rm patch}$ and the angular momentum lost by 
the patch per unit time:
\beq
\tau_{\rm patch} = -\, {d J_{\rm patch}\over{dt}} v_{\rm dr}\, {d\over{dr}}\, J_{\rm patch}(r) \, .
\label{eq-ang-momentum-evolution}
\eeq
The angular momentum of the patch is 
\beq
J_{\rm patch}(r) = M_{\rm patch} v_K r M_{\rm patch} \sqrt{GMr} \, ,
\label{eq-J_patch}
\eeq
where $M_{\rm patch}= a^2 \Sigma={\rm const}$ is the patch mass.
We thus get
\beq
\tau_{\rm patch} = {a^2 \Sigma\over 2}\, v_K v_{\rm dr}\, ,
\eeq
and by using (\ref{eq-tau_patch}), we find
\beq
v_{\rm dr} = {\gamma\over\pi}\, {rB_p^2\over{\Sigma\, v_K}} {8\gamma\over{\beta_{\rm MRI}}}\, {rP\over{\Sigma\,  v_K}}\, .
\label{eq-v_dr-1}
\eeq
Upon writing the disk surface density as $\Sigma=2\rho H$ 
introducing the isothermal sound speed $c_{\rm s}\equiv 
\sqrt{ P/\rho}$, we can rewrite $v_{\rm dr}$ as
\beq
v_{\rm dr} = {4\gamma\over{\beta_{\rm MRI}}}\, 
{r\over H}\, {c_{\rm s}^2\over{v_K}}\, .
\label{eq-v_dr-2}
\eeq
Then, using the estimate $H/r \simeq c_{\rm s}/v_K$, which follows from 
vertical hydrostatic balance in the disk, we find
\beq
v_{\rm dr} \simeq {4\gamma\over{\beta_{\rm MRI}}}\, c_s \gg 
v_{\rm accr}\simeq \alpha{H\over r}c_{\rm s} \, ,
\label{eq-v_dr-3}
\eeq
that is, the magnetized patch drifts inward with a speed of the order
of the speed of sound.

As a side remark, note that because the patches 
occupy only a minute fraction of the disk surface, their contribution 
to the overall angular momentum transport is negligible compared with 
that due to the MRI-induced turbulence in the bulk of the accretion disk. 
In fact, using equation~(\ref{eq-tau_patch}), the total magnetic torque 
per unit surface area that is attributed to patches is 
\beq
\tau = f {\gamma\over{2\pi}} rB_p^2 = 
{4\gamma\over{\beta_{\rm MRI}}}\, f r P \, ,
\label{eq-tau}
\eeq
whereas the torque per unit area due to the action of turbulence is
\beq
\tau_{\rm turb} = \alpha r P \gg \tau\, ,
\label{eq-tau-turb}
\eeq
because in this paper we assume the following ordering:
\beq
f\ll \alpha \ll \gamma = O(1) \, .
\label{eq-ordering}
\eeq


\subsection{Outward drift}
\label{subsec-outward-drift}

The inward drift just computed is partly offset by an outward drift due 
to two processes: reconnection in the disk atmosphere and a drift associated 
with the radial magnetic force. 


\subsubsection{Reconnection in the disk corona}
\label{subsubsec-recn}

The magnetic field in the disk corona is constantly changing due to the 
MRI turbulence inside the disk. Because of the high Alfv\'en speed in 
the corona, the external open field lines may rapidly reconnect to this 
changing field. This causes an effective diffusion of the external field
through the disk. 
If $L\simeq H$ is the radial length scale of the MRI turbulence, and 
$\epsilon\Omega$ is the rate at which the coronal field changes due to 
the turbulence in the disk, reconnection to this changing field causes 
diffusion with a coefficient that can be estimated as
\beq 
D={1\over 3}L^2\epsilon\Omega\simeq {\epsilon\over 3}\Omega H^2 \, .
\eeq
We can estimate $\epsilon$ from the results of MRI simulations. 
These produce a field in which the azimuthal component dominates, 
$B_\phi/B_r\sim 10-20$ (e.g. Brandenburg et al. 1995). This azimuthal 
field is created by stretching of a radial component in the Keplerian 
shear $\sigma={3\over 2}\Omega$, which takes a time $B_\phi/(\sigma B_r)$. 
Thus, the rate at which the radial field component changes must be of the 
order $\sigma B_r/B_\phi$, hence $\epsilon={3\over 2}B_r/B_\phi\simeq 0.1$, 
or $D\simeq 0.03 H^2\Omega$. This is of the same order as the disk viscosity 
$\nu=\alpha c_{\rm s}^2/\Omega$,
\beq 
D/\nu={\epsilon\over{3\alpha}} \simeq {0.03\over\alpha} \, .
\label{dnu}
\eeq
The effect of this diffusivity by reconnection in the disk corona is the same 
as magnetic diffusion across the disk, and if it were the only effect present 
it would limit the angle at which an external field can be dragged in to 
$\simeq H/r\ll 1$, as before.  Equivalently it would cause an outward drift 
at a speed of the order $v=\epsilon c_{\rm s}/3$.


\subsubsection{Viscous drift}
\label{subsubsec-viscous-drift}

If $B_r=k B_p$ is the radial component of the field of a patch (or rather, 
the part of it on which the magneto-centrifugal mass loss takes place) just
above the disk surface, the outward force on a patch of area $a^2$ is 
\beq 
F_r= 2 a^2 k B_p^2/(4\pi) \, .
\eeq
To lowest order, this force is balanced by a small deviation from Keplerian 
rotation, so that the patch moves relative to its surroundings with velocity
\beq 
\delta v_{\phi} = kv_{\rm A}^2/{2c_{\rm s}} \, . 
\eeq
This difference implies a frictional force on the patch in the azimuthal 
direction, which causes a gradual change in angular momentum, resulting 
in an outward drift. The friction can be computed either as a turbulent 
drag or a viscous drag based on the assumed turbulence in the disk. 
We assume here a viscous drag, turbulent drag would give qualitatively 
similar results. The drag force on a magnetic patch of of area $a^2$ 
can then be written as $C \rho\nu a \delta v_{\phi}$, where 
$\nu=\alpha c_{\rm s}^2/\Omega$ is the disk viscosity and 
$C$ is a constant numerical factor.


\subsubsection{Net drift speed}
\label{subsubsec-net-drift}

The rate of angular momentum change by the combined effects of magnetic 
angular momentum loss and drag as estimated above is then
\beq 
\dot J_{\rm patch} = {Ck\alpha\over 2} raH\, {B_p^2\over{4\pi}} - 
\gamma ra^2\, {B_p^2\over{2\pi}} \, .
\eeq
This corresponds to a radial drift speed 
\beq 
v_{\rm d}={\dot{J}_{\rm patch}\over{{\rm d}J_{\rm patch}/{\rm d}r}} {2v_{\rm A}^2\over c_{\rm s}}\, 
\biggl( {C\alpha kH\over{4a}} - \gamma\biggr)\, ,
\eeq
where the patch angular momentum $J_{\rm patch}$ is given by 
equation~(\ref{eq-J_patch}). To compute the net inward drift, 
the effect of reconnection in the atmosphere must be added. 
If the tangent of the angle between the external poloidal 
field and the vertical is $B_r/B_z=k$, then the effective 
reconnection of $B_r$ across the disk causes the vertical 
field component to drift at a rate $kD/H$, with the diffusion 
coefficient $D$ given by equation~(\ref{dnu}). Adding this to 
the above gives the net drift speed
\beq  
v_{\rm d}={2v_{\rm A}^2\over c_{\rm s}}\, (C\alpha kH/4a-\gamma)\, + 
\, {k \epsilon\over 3}\, c_{\rm s} \, .
\eeq
For a stationary state $v_{\rm d}=0$, 
\beq 
k= {B_r\over B_z} = {6\over\epsilon}{v_{\rm A}^2\over c_{\rm s}^2}\, 
(\gamma-C\alpha kH/4a)\, .
\label{angle}
\eeq
For inward drift, $B_r/B_z>0$, we must have
\beq 
\gamma>\alpha kH/a \, , 
\label{requ}
\eeq
that is, the angular momentum loss in the patches (measured by $\gamma$) 
must be large enough to overcome their outward viscous drift. If this is 
satisfied, the magnetic field is dragged in to the angle given by~(\ref
{angle}), i.e., $B_r/B_z\simeq (6\gamma/\epsilon)\,v_{\rm A}^2/c_{\rm s}^2$. 

The significance of this result is that the angle $B_r/B_z$ can be of 
{\it order unity}, compared with the turbulent diffusion model of external 
fields in a disk where it is only of order $H/r$. The requirement~(\ref{requ}) 
still has to be satisfied, of course, and whether this is realistic will need 
further investigation.

To illustrate what this implies for the degree of inward concentration of 
the field that can be achieved, note that a potential field matching to a 
disk with a vertical magnetic field varying as $B_z\sim r^{-\mu}$ has a 
constant angle at the disk surface (Sakurai 1987):
\beq
B_r/B_z={\Gamma({\mu + 1\over 2})\Gamma(1-{\mu-1\over 2}) \over 
\Gamma({\mu\over 2}) \Gamma(1-{\mu\over 2})} \, .
\eeq
This is an increasing function of $\mu$, and an angle of 45$^\circ$ 
for example, corresponds to $\mu=1$.
In an X-ray binary disk with outer radius $R_{\rm o}=10^{11}$ cm and 
inner edge $r_{\rm i}=10^7$ cm, for example, this would yield field 
strength ratios $B_{\rm i}/B_{\rm o}\simeq 10^4$ for $\mu=1$.


\subsection{Global Evolution of the Large-Scale Magnetic Field}
\label{subsec-global-evoln}

Here we determine how the radial drift of individual patches
of strong magnetic field contributes to the evolution of the  
disk's poloidal flux distribution on the global scale. 
In order to do this, we now consider all the local quantities discussed 
in the preceding section (e.g., $B_p$, $f$, $B_0$, $v_{\rm dr}$) to be 
functions of two variables --- time $t$ and cylindrical radius $r$. 
Then, the continuity equation for $\Psi(r,t)=\int\limits_0^r B_0(r,t) 
2\pi r dr$  --- the vertical magnetic flux threading the disk inside 
the radius $r$ --- can be written as
\beq
{\partial\Psi\over{\partial t}} = 2 \pi r B_0(r,t) v_{\rm dr}(r) \, .
\label{eq-dPsidt}
\eeq
Correspondingly, the equation of evolution for the large-scale
vertical magnetic field $B_0(r,t)$ is
\beq
{\partial B_0\over{\partial t}} = 
{1\over r}\, \partial_r \, [r B_0(r,t) v_{\rm dr}(r)] \, .
\label{eq-dB0dt-1}
\eeq
As we look at the time evolution of the magnetic flux
we assume that it takes place on timescales much shorter
than the overall accretion timescale, so that the parameters
describing the properties of the disk, such as $M$ and 
$\dot{M}$, etc., can be taken as constant in time.

Since $v_{\rm dr}$ is independent of $B_0$, equation~(\ref{eq-dB0dt-1}) 
is linear in $B_0$ and so can be easily analyzed. For $v_{\rm dr}$ given 
by expression (\ref{eq-v_dr-3}), we get:
\beq
{\partial B_0\over{\partial t}} = {4\gamma\over\beta_{\rm MRI}}\, 
{1\over r}\,  \partial_r [r c_s(r) B_0(r,t)] \, .\label{eq-dB0dt-2}
\eeq
where we have assumed $\gamma$ and $\beta_{\rm MRI}$ to be constant. 

For example, for a standard Shakura--Sunyaev disk model in 
the gas-pressure dominated regime with Thomson opacity (e.g.,\ 
Frank et al. 2002) $H(r)\sim r^{21/20}$, i.e., the disk vertical 
scale height $H$ scales almost linearly with radius at large distances 
($r\gg r_g$). Thus we shall regard the ratio 
\beq
{H\over r} \simeq {c_s\over{v_k}} \ll 1 
\label{eq-def-epsilon}
\eeq
as a small constant parameter in our model. Then we can write
\beq
{\partial B_0\over{\partial t}} = {4\gamma\over {\beta_{\rm MRI}}}\, 
{H\over r}\, {\sqrt{GM}\over r}\, \partial_r [\sqrt{r} B_0(r,t)] \, .
\label{eq-dB0dt-3}
\eeq

So far we assumed that the large-scale transport of the vertical
magnetic flux is determined solely by an uninhibited inward drift
of magnetic patches in accordance with formula (\ref{eq-dB0dt-1}). 
This corresponds to the assumption that there is an effective sink 
for the flux somewhere at small radii. Although this is not realistic, 
this situation is a fair approximation for what happens at large 
distances and at early times.

In particular, if we consider some intermediate range of radii in the 
middle of the disk (i.e., sufficiently far away from both the inner 
and outer disk edges), it follows from equation~(\ref{eq-dB0dt-3}) 
that the steady state distribution of the net vertical field is:
\beq
B_0(r)\sim 1/\sqrt{r} \, .
\label{eq-stationary}
\eeq

At the same time, the magnetic field $B_p$ inside the patches
just scales as the square root of the gas pressure. In the case 
of a radially self-similar disk, we have [e.g., eqn (7.31) in 
Krolik 1999]:
\beq
\int p(z)\, dz = H P \sim r^{-3/2}\, ,
\eeq
and since we here take $H \sim r$, we get $P(r)\sim r^{-5/2}$,
and hence $B_p\sim r^{-5/4}$. Therefore, the covering fraction
of our patches scales with radius as 
\beq
f\sim B_0(r)/B_p(r) \sim r^{3/4} \, ,
\eeq
i.e. $f$ increases outward.


\section{The central flux bundle}
\label{sec-bundle}

\begin{figure}
\plotone{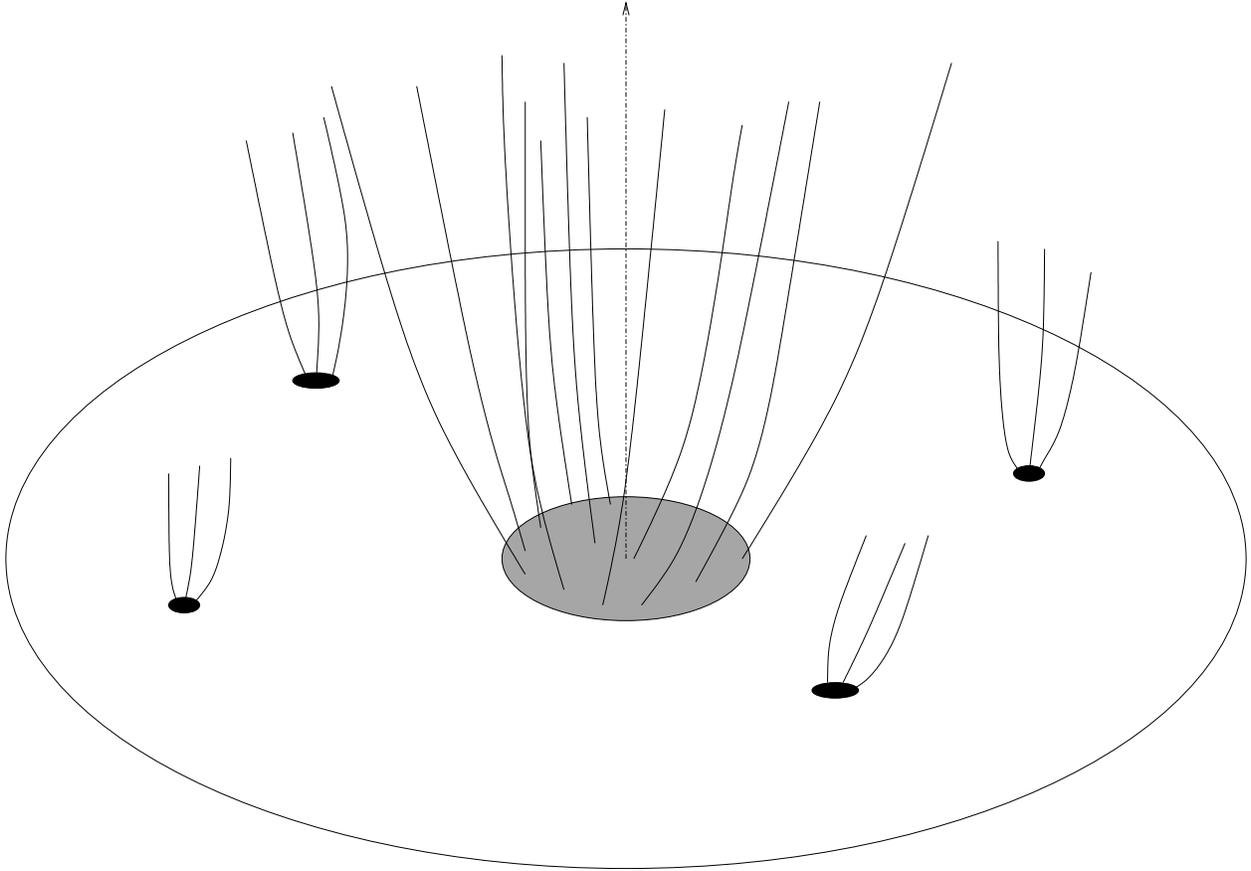}
\caption{A large central magnetic flux bundle (gray) develops as a result 
of continuing accumulation of many small strongly-magnetized patches (black)
drifting inward through the disk.\label{fig-2}}
\end{figure}

The inward drift of the patches is halted by the central object, 
so that a central bundle (called a Magnetically Arrested Disk by 
Narayan et al. 2003) of accumulating flux develops around it
(see Fig.~\ref{fig-2}). In a steady state, mass continues to 
accrete through this bundle, but, due to the high field strength, 
the processes that mediate the accretion flow are now different 
from those in a normal accretion disk. For example, Narayan et al. (2003)
proposed that accretion takes place in the form of blobs and streams 
(see their Fig.~1).


\subsection{Accretion through the central bundle}

The strong field suppresses MRI turbulence so that the effective viscosity 
is reduced. The resulting pile-up of mass outside of the bundle eventually 
becomes unstable to interchange instabilities at the bundle's outer edge. 
In this way mass enters the bundle while magnetic flux from the bundle 
mixes outward into the disk. These instabilities have been studied by 
analytical means (in the WKB approximation) by Spruit et al. (1995) and 
by Lubow \& Spruit (1995). The onset of small-scale modes typical of 
interchanges (as in Rayleigh-Taylor) takes place only at rather large 
field strengths, due to a stabilizing effect of the Keplerian shear. 

Numerical simulations of a strong ordered vertical field in a disk 
(Stehle 1996; Stehle \& Spruit 2001) show the behavior typical of 
interchange instability, but only at low shear rates (less than Keplerian). 
In Keplerian shear, on the other hand, a large-scale, a global (azimuthal 
order $m=1$) mode sets in first, and dominates the nonlinear development 
of instability (see also Caunt \& Tagger 2001). The simulations show that 
the net effect of this global instability, however, is similar to that of 
an interchange: the strong field spreads outward while mass accretes inward. 
This scenario agrees well with the recent 3D MHD simulations by Igumenshchev
et al. (2003) who have observed radiatively-inefficient accretion in the 
form of blobs and streams produced by interchanges.

The field strength at which this instability becomes effective is most 
usefully expressed in terms of the degree of support against gravity 
provided by the magnetic stress~$B_rB_z$. The simulations indicate 
that the instability becomes effective when the radial acceleration due to 
the magnetic force is of the order of a few per cent of the gravitational
acceleration. In a cool disk, this corresponds to a field strength 
that is large compared with the gas pressure in an otherwise identical 
non-magnetic disk. 
Thus, as long as the field line inclination $B_r/B_z$ is of order unity, 
there is a range in field strengths, between the value at which MRI 
turbulence is suppressed and the value where dynamical instability 
of the bundle itself sets in, where no known instability operates 
(Stehle \& Spruit 2001). In this range, no accretion can take place.
Instead, mass would build up outside a region with such field strengths 
until the central bundle is compressed enough for instability to set in. 
At the edge of the bundle, where the field is more horizontal, $B_r/B_z$ 
is larger and instability is expected already at lower values of the vertical 
component $B_z$ (see illustration in Fig.\ \ref{fig-bundle}).

In a steady state the flux bundle stays in a state somewhat above marginal 
stability. The instability settles at such an amplitude that it allows mass 
to accrete at the incoming rate, across the magnetic field and onto the black 
hole. 




\subsubsection{A feedback effect}

The central flux bundle has an `attractive' effect on nearby flux patches 
that have not yet merged with it, due to a slightly counterintuitive effect.
The field of the central bundle fans out above the disk, so that
its field lines skim the disk surface. This strong radial field 
affects nearby patches by pushing their field lines over in the 
radial direction as well (see Fig.~\ref{fig-bundle}). This makes 
the conditions for magneto-centrifugal mass loss from the patches 
more favorable. The consequent angular momentum loss increases the 
inward drift speed of the patches, as if they were attracted towards 
the central bundle.


\subsection{Transition between central bundle and accretion disk}

\begin{figure}
\plotone{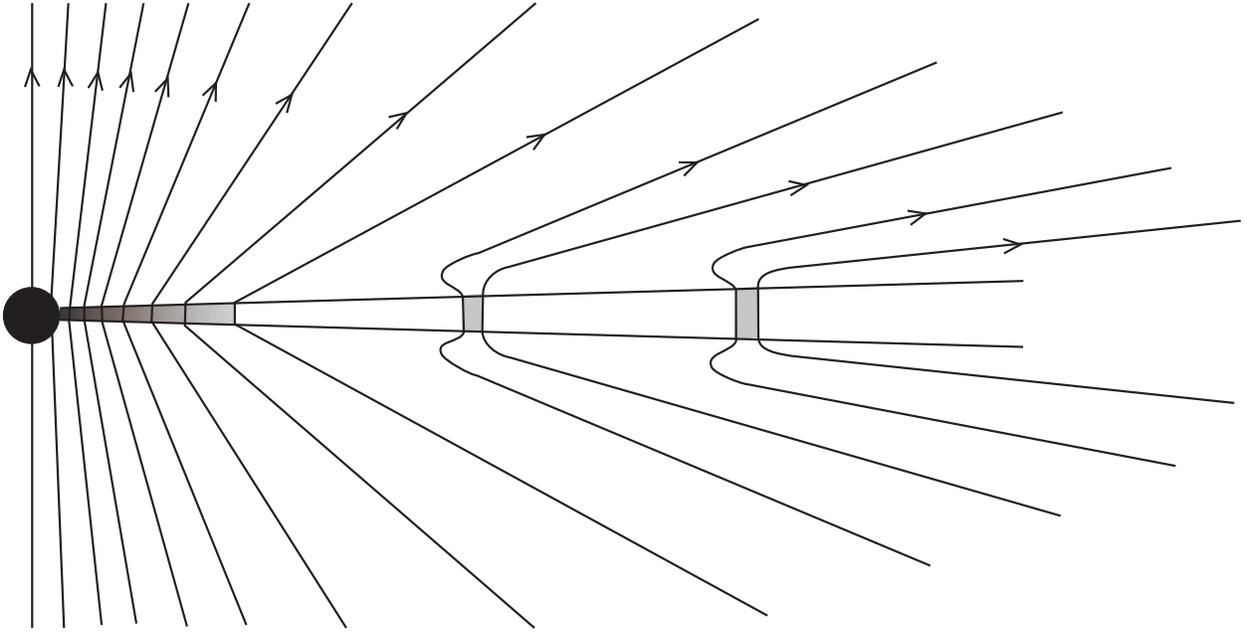}
\caption{Accretion disk with magnetic patches approaching a 
central flux bundle (schematic). The figure shows the open field 
lines (connecting to infinity). Not shown is the turbulent small 
scale field in the disk itself. Shading indicates the strength of 
the open field lines at the points where they pass through the disk. \label{fig-bundle}}
\end{figure}

The radial acceleration by the magnetic field of the central bundle, on 
a thin disk of surface density $\Sigma$ embedded in it is
\beq
g_{\rm m}=B_rB_z/(2\pi\Sigma).
\eeq
If at the onset of internal instabilities in the bundle this
acceleration is a fraction~$\epsilon$ of the acceleration of gravity, as
described above, the magnetic field $B_z$ at that point is larger than
equipartition with the gas pressure by a factor of order
$\sqrt{(\epsilon/k)(r/H)}$, where $k$ is the ratio $B_r/B_z$ at the disk
surface:
\beq 
B_z\approx B_{\rm eq}\sqrt{(\epsilon/k)(r/H)} \, .
\label{eq-transition}
\eeq
We can use this to estimate how the transition from a field consisting
of inward moving patches to the field of the central bundle takes place. 
Since the filling factor of the strong-field patches is low outside the 
central bundle, the inclination $k$ of the magnetic field in the bundle 
is high near its edge (Fig.~\ref{fig-bundle}), and $B_z$ is correspondingly 
low [see eq.~(\ref{eq-transition})]. 
The vertical field strength in the bundle thus decreases smoothly outward 
from a high value at the center to a value near $B_{\rm eq}$ at its edge. 
Near this point it matches the value of the  magnetic patches approaching 
the bundle, which have field strength of order equipartition with the gas
pressure (section~\ref{subsec-enhan}). From equation~(\ref{eq-transition}), 
the inclination of the field at this point is of the order $k\simeq\epsilon 
r/H$. This is sketched qualitatively in Figure~\ref{fig-bundle}.


\subsection{The size of the central bundle}

The flux in the central bundle increases with time, as long as the sign 
of the external flux captured by the disk stays constant. When the sign 
of the accreted flux changes, all the processes described above remain 
unchanged, except that the flux in the central bundle starts to decrease. 
Thus, the amount of flux in the bundle reflects the history of the field 
trapped by the disk from its environment. Its variations with time will 
reflect conditions near the outer edge of the disk. These are likely to 
be very long compared with accretion time scales near the hole.


\section{Discussion and conclusions}
\label{sec-conclusions}

The capture of external magnetic flux by an accretion disk and its subsequent 
compression in the inner regions of the disk is an attractive possibility to 
explain both the missing `second parameter' determining the X-ray spectrum 
and jet activity. In this paper, we have argued that this process can actually 
be more relevant than it is usually assumed, provided that the external field 
can be concentrated  into patches of field comparable in strength to the 
magnetorotational turbulence in the disk. 

Because the external field lines are open, they can carry a magnetically 
driven wind. The angular momentum loss through this wind is transmitted 
to the disk by the footpoints of these external field lines. As a result, 
these are the first parts on the disk to move inward, carrying the external 
field lines with them. At the same time, the patchiness of these field lines 
at the disk surface guarantees that conditions favorable for launching a wind 
exist on a significant fraction of them. 

We find that the inward drift speed of the external field lines can reach 
a significant fraction of the sound speed, i.e., a factor $r/(\alpha H)$ 
faster than the viscous accretion velocity.

As a result of this inward drift of strongly magnetized patches, 
the external field lines accumulate at the center of the disk into 
a flux bundle. The poloidal flux of this bundle grows as long as 
the polarity of the external flux trapped by the disk does not change. 
As discussed in the Introduction, there is now extensive observational 
evidence that the accretion flow in X-ray binaries does not depend solely
on the instantaneous mass accretion rate in the inner disk; instead, a 
`second parameter' must also be involved. We suggest that the size of 
the central magnetic flux bundle is to be identified with this second 
parameter.

For accretion to proceed, there must be processes allowing mass to drift across the central flux bundle. These will be different from those acting in a normal accretion disk. Since the central flux bundle is also the region where most of the gravitational energy is released from the accreting mass, the strong variability observed in the so-called hard states of black hole and neutron star accreters may also be a consequence of the processes specific to  mass transfer across a central bundle of magnetic flux. 

The arguments given here for the processes involved in the capture and inward 
drift of an external field are admittedly qualitative, but they may be tested 
to some extent by numerical simulations. Because of the large Alfv\'en speeds 
encountered in the disk atmosphere, and the need to resolve small scales in 
the disk as well as larger scales in the captured flux, these will be somewhat 
challenging, however, unless a way is found to use complementary approximations for the disk and its atmosphere. A useful first step in this direction has been made by Stehle \& Spruit (2001), where the existence of instabilities 
that allow for mass to accrete across the central flux bundle was demonstrated.

One of us (D.U.) expresses his gratitude to Max-Planck-Institut 
f\"ur Astrophysik (MPA) for its warm hospitality during D.U.'s 
visit to MPA in Spring 2004, during which this work has originated. 
This research has been supported in part by the National Science 
Foundation under Grants Nos.~PHY99-07949 (KITP) and~PHY-0215581
(PFC: Center for Magnetic Self-Organization in Laboratory and 
Astrophysical Plasmas).


\appendix
\section{Angular momentum loss as a function of the outflow mass flux}

In the cold magnetocentrifugal wind model, the angular momentum loss per
unit of poloidal magnetic flux is
\beq 
j=\dot m\Omega\varpi_{\rm A}^2 \, ,
\eeq
where the mass flux per unit of poloidal magnetic flux is $\dot m=\rho
v_{\rm p}/B_{\rm p}$ (which is constant along a field line), $v_{\rm p}$
and $B_{\rm p}$ are the poloidal velocity and field strength, and 
$\varpi_{\rm A}$ the distance of the Alfv\'en surface from the rotation
axis. The solution for a cold wind in a radial poloidal field yields 
(e.g., Spruit 1996):
\beq 
j={B_0\over 4\pi}\, \varpi_0 f(\mu), ,
\eeq
where $B_0,\varpi_0$ are the poloidal field strength and distance from
the axis at the base of the flow, and the dimensionless function~$f$ is
\beq 
f(\mu)=\, {3\over 2}\, \mu (1+\mu^{-2/3}) \, , 
\label{eq-fmu} 
\eeq
where $\mu$ is a dimensionless mass flux:
\beq 
\mu=4\pi \dot m \Omega\varpi_0/B_0 \, .
\eeq

At the base of the flow, where $v_{\rm p}=v_{{\rm p}0}$, $\rho=\rho_0$,
this can be written as
\beq
\mu=\rho_0v_{{\rm p}0}\, 4\pi\Omega\varpi_0/B_0^2=v_{{\rm p}0}
\Omega\varpi_0/v_{{\rm A}0}^2 \, ,
\eeq
where $v_{{\rm A}0}$ is the poloidal Alfv\'en speed at the base.

Per unit of surface area, the angular momentum flux is $\dot J=B_{\rm p}j$, 
which can be written as
\beq
\dot J={B_0^2\over 4\pi}\varpi_0f(\mu).
\eeq
While this result has been derived for the specific geometry of a radial
poloidal field (the `split monopole'), numerical solutions show that it
also holds well for more general geometries (Anderson et al. 2004).

Comparing this with the angular momentum flux in equation~(\ref{eq-tau_patch})
(and noting that in this equation the sum has been taken over the two 
surfaces of the disk), we find that
\beq 
\gamma=f(\mu) \, .
\eeq

Thus, $\gamma \gg 1$ if $\mu\gg 1$ and $\gamma \ll 1$ if $\mu\ll 1$.



\clearpage


\clearpage

\clearpage

\end{document}